# Studies of the temperature and frequency dependent impedance of an electroceramic functional oxide thermistor

*Rainer Schmidt[1,2], Andrew W. Brinkman[1]

[1]Department of Physics, University of Durham, South Road, Durham DH1 3LE, United Kingdom

[2] now at: Department of Materials Science & Metallurgy, University of Cambridge, Pembroke Street, Cambridge CB2 3QZ, United Kingdom;

*corresponding author's e-mail: rs441@cam.ac.uk, Fax: 0044 1223 334373

## Abstract

The charge transport mechanism and the macroscopic dielectric constant in polycrystalline device materials commonly exhibit several components such as electrode-sample interface, grain boundary and bulk contributions. In order to gain precise understanding of the functionality of polycrystalline electroceramic device materials it is essential to deconvolute these contributions. The paradigm of functional thermistor ceramics based on thick film spinel manganates has been studied by temperature dependent alternating current impedance spectroscopy. Three typical relaxation phenomena were detected, which all showed a separated temperature dependence of resistivity consistent with thermally activated charge transport. The dominating grain boundary and the interface contributions exhibited distinctively different capacitance allowing clear identification. The composite nature of the dielectric properties in polycrystalline functional ceramics was emphasized, and impedance spectroscopy was shown to be a powerful tool to account for and model such behaviour.

# 1. Introduction

Spinel type manganates are high technology electroceramic materials, which are commonly used in a wide range of thermistor and related applications.[1] Historically, the compound NiMn$_2$O$_4$ has been the basis of NTCR thermistor developments,[2] but nowadays a wide range of dopants are commonly included into the system to increase stability,[3] reduce ageing effects[4] or tune the NTCR thermistor parameters.[5]

The key material property relevant for thermistor applications is the resistivity vs temperature ($\rho$-$T$) behaviour. Intrinsic charge transport is thought to be by localised electron hopping and the material exhibits a uniform negative temperature coefficient of resistance (NTCR) between 140 K – 520 K without any indication of electronic phase transitions.[2] In polycrystalline materials the resistive and dielectric properties usually consist of several components, such as electrode - sample interface, grain boundary and intrinsic bulk contributions, which can be separated and quantitatively determined in terms of their resistivity and capacity by alternating current (ac) Impedance Spectroscopy (IS) measurements.[6] Solid state IS on ceramic materials is commonly employed to (a) deconvolute different contributions in ionic conductors such as yttria and samaria doped ceria,[7] (b) to detect mixed ionic–electronic charge transport for example in yttria stabalised zirconia[8], or (c) to characterise insulating materials such as dielectrics[9] and ferroelectrics.[10,11] The method has been scarcely employed though to analyse electrically conducting ceramics;[12] studies on PTCR thermistors have been published before,[13] but no comprehensive IS characterisation of NTCR thermistors is available at present.

In the temperature range relevant for NTCR thermistor applications the direct current (dc) $\rho$-$T$ curve of spinel nickel manganates has been shown to follow a small-polaron hopping model in the non-adiabatic regime for Nearest-Neighbour- or Variable-Range- Hopping (NNH, VRH),[14] typical for transition metal oxides with a strongly localised character of electron charge carriers. Both types of hopping can be described by a generalised expression:[15]

$$\rho(T) = D\, T^{\alpha} \exp\left(\frac{T_0}{T}\right)^{p} \tag{1}$$

$D$ is the temperature independent contribution to the resistivity, $T_0$ a characteristic temperature, $\alpha$ describes the pre-exponential temperature dependence, and $p$ the exponential power law. For NNH $p = \alpha = 1$ and for conventional VRH $0.25 < p = \alpha/2 < 1$.[16] It was shown by dc $\rho$ vs $T$ measurements that the macroscopic hopping exponent $p$ varies in different types of spinel nickel manganate films with different microstructure.[14] Here, the intrinsic $\rho$-$T$ and $p$ behaviour of thick films for each resistive component is investigated separately and its influence on the macroscopic dc resistivity is studied. For thermistor and related applications it is important to develop an accurate understanding of the composite nature of the $\rho$-$T$ characteristic as the essential feature of the device mechanism. Fitting the frequency dependent specific impedance of a thermistor to an appropriate equivalent circuit model at various temperatures allows obtaining and plotting the resistivity and capacitance of each contribution vs $T$.

## 2. Solid State Impedance Spectroscopy

Ac impedance spectroscopy experiments consist effectively of a time ($t$) dependent alternating voltage signal of angular frequency $\omega$ with a shape of $U(\omega,t) = U_0\cos(\omega t)$ applied to a sample, and the phase shifted current response signal is measured: $I(\omega,t) = I_0 \cos(\omega t - \delta)$. One phase of the applied voltage signal corresponds to a $2\pi$ rotation of the voltage arrow $U(\omega,t)$ in Fig 1. The current response of ideal equivalent circuit elements are (1) in phase with the applied voltage in case of a resistor R, (2) out of phase by $\delta = -\pi/2$ for a capacitor C and (3) out of phase by $\delta = +\pi/2$ for an inductor.

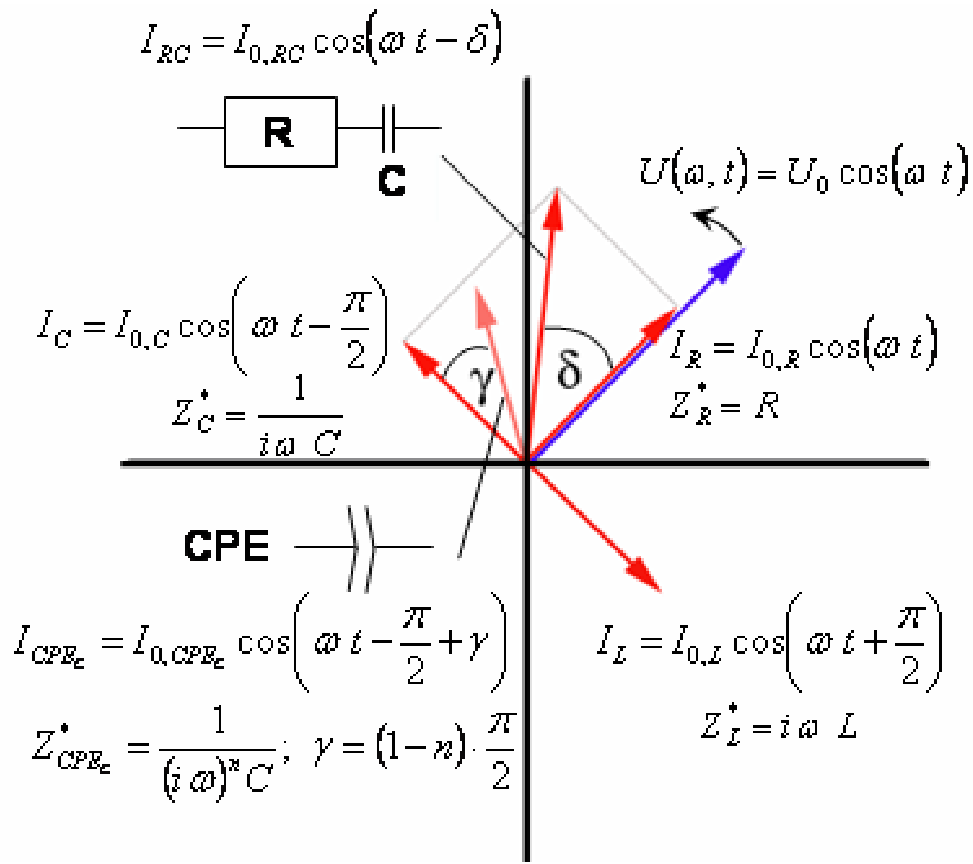

**Fig 1** Impedance response of model equivalent circuit elements on a phase arrow diagram: applied voltage $U(\omega t)$, current response $I_R$ for an ideal resistor R, $I_C$ for an ideal capacitor C, $I_{RC}$ for a series resistor-capacitor combination RC, $I_L$ for an inductor L, and $I_{CPE/C}$ for a constant phase element $CPE_C$

All phase angles are time independent for a given frequency, which allows defining the impedance as a time independent complex number. The impedance fraction in phase with $U(\omega,t)$ is defined as the real, and the part +/- $\pi/2$ out of phase as the positive/negative imaginary part. The complex definitions of the impedance $Z^*$ of all model circuit elements are given in Fig 1. Impedance spectra of separated real and imaginary parts are commonly obtained by measuring $I_0$ and $\delta$ over a large frequency range (typically $f \sim$ 10 Hz – 2 MHz). The complex impedance of electrode - sample interface, grain boundary (GB) and bulk relaxation phenomena in polycrystalline materials can each be described by a RC element consisting of a resistor and capacitor in parallel, where the macroscopic impedance is just the sum of the individual RC impedances (see Fig.2).[6] According to the classification scheme proposed in ref. [6], the magnitude of each specific capacitance $c$ (= $\varepsilon_0 \cdot \varepsilon'$) allows the originating effect of the relaxation to be identified (electrodes, GB or bulk). $\varepsilon'$ is the real part of the dielectric constant of the respective capacitor and $\varepsilon_0$ for vacuum.

In order to fit real system impedances, R and C can be replaced by Constant Phase Elements (CPE) to account for non-Debye behaviour. A CPE exhibits a frequency independent constant phase angle $\gamma$ with respect to the response from an ideal capacitor (Fig 1) or resistor. The specific impedances [Ohm·cm] of CPEs describing real capacitors (CPE$_C$) and resistors (CPE$_R$) are given by:

$$z^*_{CPE_C} = \frac{1}{c^m (i\omega)^n} \; ; \; z^*_{CPE_R} = \frac{\rho_{dc}^m}{(i\omega)^q} \tag{2}$$

where $c^m$ is the specific capacitance in modified units of $F \cdot s^{n-1} \cdot cm^{-1}$, $\rho_{dc}{}^m$ is the resistivity in modified units of $\Omega \cdot s^{-q} \cdot cm$; $n \leq \sim 1$ and $q \geq \sim 0$. The phase shift with respect to the applied voltage signal is $\delta = -\pi/2 \cdot n$ and $\delta = -\pi/2 \cdot q$. Usually, an efficient fit of the sample impedance to an equivalent circuit can not be obtained using multiple $CPE_R$-$CPE_C$ elements, because 4 free parameters for each $CPE_R$-$CPE_C$ lead to an over-determined model. For electroceramic sample response presented in the impedance notation, usually the non-Debye contribution of the resistance is neglected, and the equivalent circuit is made up of R-$CPE_C$ (ZARC) elements with a specific impedance of:[17]

$$z^*_{R-CPE_C} = \frac{\rho_{dc}}{1 + \rho_{dc} \, c^m \, (i\omega)^n} \qquad (3)$$

The reasons for non-Debye response are difficult to determine, and a variety of causes have been suggested.[4] In stable ceramic polycrystalline materials a distribution of relaxation times $\Delta \tau$ ($\tau = \rho_{dc} \cdot c$) in the sample may be the most plausible explanation.[18] In fact, the parameters $\rho_{dc}$ and $c$ may both display independent distributions. For electrode sample interface effects $CPE_C$ behaviour has been associated alternatively with the fractal behaviour of the electrode sample interface with a dimensionality $d$ of $2 < d < 3$, and $d$ can be directly related to $n$.[19]

Impedance spectroscopy data is commonly plotted as a Nyquist locus of specific (or absolute) negative imaginary vs real part of the impedance $-z''$ vs $z'$ (or $-Z''$ vs $Z'$), where in the ideal Debye case each RC element is represented by a semicircle of radius $\rho_{dc}/2$ ($R/2$) with a maximum in $-z''$ ($-Z''$) at

$$\omega_{max} = \frac{1}{RC} = \frac{1}{\tau} = \frac{1}{\rho_{dc} \, c} \tag{4}$$

A non-ideal R-CPE$_C$ element yields a suppressed semicircle. In the case of RC/ R-CPE$_C$ elements with variation in time constant $\tau$ of at least 3 orders of magnitude, semicircles are fully developed with marginal overlap. In experiments strong overlap is often observed and numerical fitting tools are needed to determine reliable $\rho$ and $c$ values. From the frequency dependent complex specific impedance $z^*$ ($= z'- iz''$) other parameters can be derived, such as the complex dielectric constant $\varepsilon^*$ ($= \varepsilon'- i\varepsilon''$), capacitance $C^*$ ($= C'- iC''$), conductance $Y^*$ ($= Y'+ iY''$) and modulus function $M^*$ ($= M'+ iM''$).[17]

$$z^* = \frac{1}{i\omega \, \varepsilon_0 \, \varepsilon^*} = \frac{g}{Y^*} = \frac{M^*}{i\omega \, \varepsilon_0}; \quad \varepsilon^* = \frac{C^*}{C_0} = \frac{c^*}{\varepsilon_0} \tag{5}$$

where $C_0$ is the capacity of the measurement cell in vacuum, i.e. $C_0 = \varepsilon_0 \cdot A/d$. $A$ is the current cross section and $d$ the distance of the contacts. $A/d$ is referred to as the geometrical factor $g$ [cm].

## 3. Results and Discussion

In a previous publication, the Nyquist loci of raw data taken from a thick film nickel manganate thermistor have been presented,[20] where a single Debye semicircle was aligned with the Nyquist loci and a resistance vs temperature curve was obtained from the semicircle dimensions at different temperatures. Here it is shown that this approach is insufficient to obtain a viable fit to the data, and additional contributions at the low and high frequency ends need to be taken into account. Furthermore, the raw data was normalised here by the appropriate parameter to allow a valid classification of the relaxation by determining $c$. The following equivalent circuit has been used for data fitting using commercial software (Z-View 2):

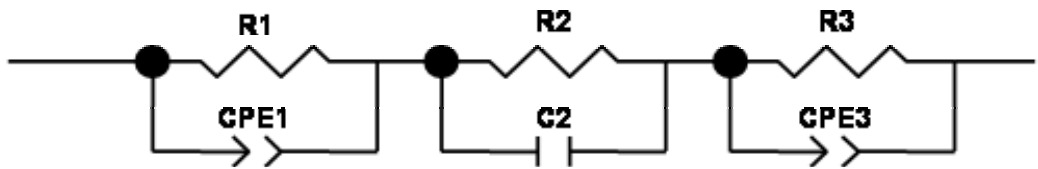

**Fig 2** Equivalent circuit model

The main impedance contribution from R2-C2 was assumed to exhibit ideal Debye response, which was justified by a near-ideal semicircle dominating the Nyquist loci. Marginal deviations of R2-C2 from Debye response may have affected the fits of R1-CPE1 and R3-CPE3. Therefore, quantitative analysis of the parameters of both R-CPEs may be regarded with care due to extrinsic influences. At low temperatures the R3-CPE3 element was evident in the spectra and its use justified, but it needed to be omitted from the circuit at 120°C and above. Still, at $T \geq 120°C$ the high frequency data was slightly affected by R3-CPE3 and was therefore cut off at lower $f$. A further restriction was made by setting a fixed $n$-value for CPE1 ($n1$). For a free $n1$ parameter a poor fit was obtained, because the model was over-determined due to the limited data available in the low $f$ regime. In Fig 3(a) the advantage of a fit using a fixed over a free parameter $n1$ is demonstrated, and is compared to a fit for one single Debye RC element.

A frequency independent specific capacitance $c'$, expected for a single RC, is clearly not in agreement with the data. The frequency dependent capacitance in the low $f$ regime may represent a crossover region of the impedance dominated by R1-CPE1 to domination by R2-C2. The full fit at various temperatures using the Fig 2 equivalent circuit model with fixed $n1$ is displayed in Fig 3(b), plotted as Nyquist loci. Fig 3(b) shows that the data was described satisfactorily at all temperatures, only at the low frequency ends the model deviated marginally. Semicircles are distorted due to the logarithmic axes.

**(a)**

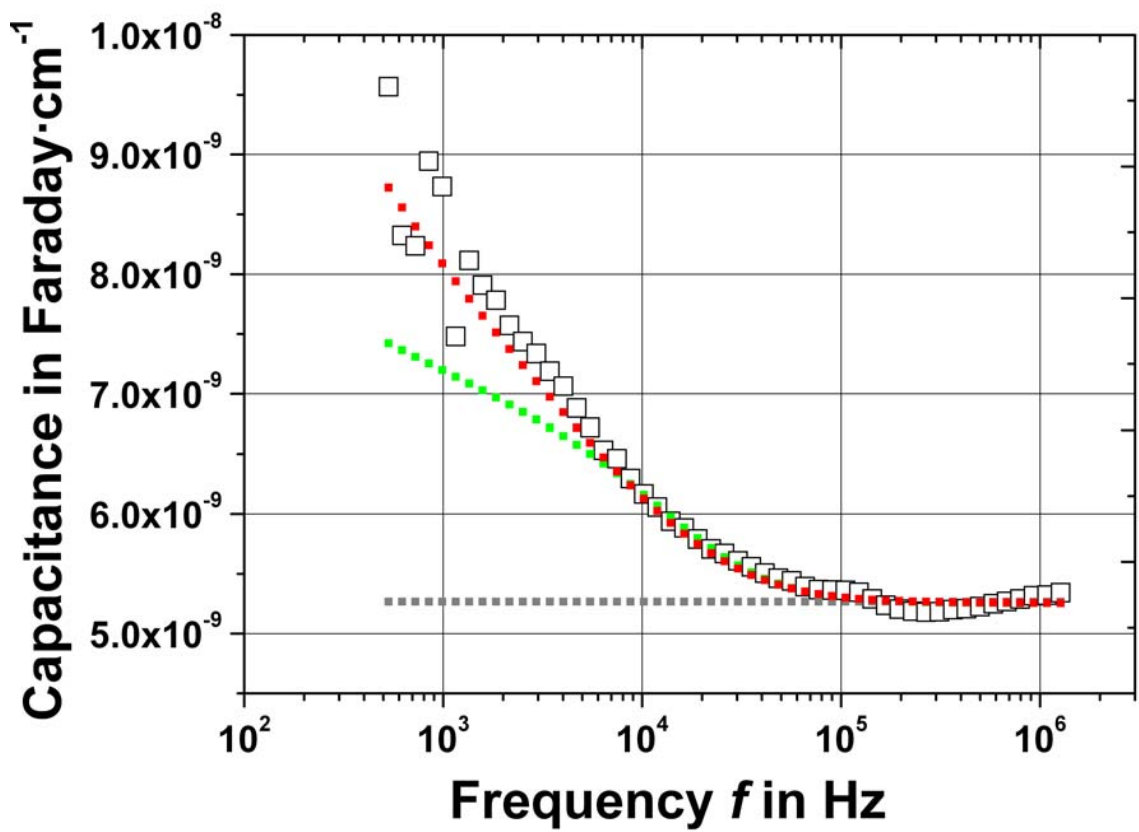

**(b)**

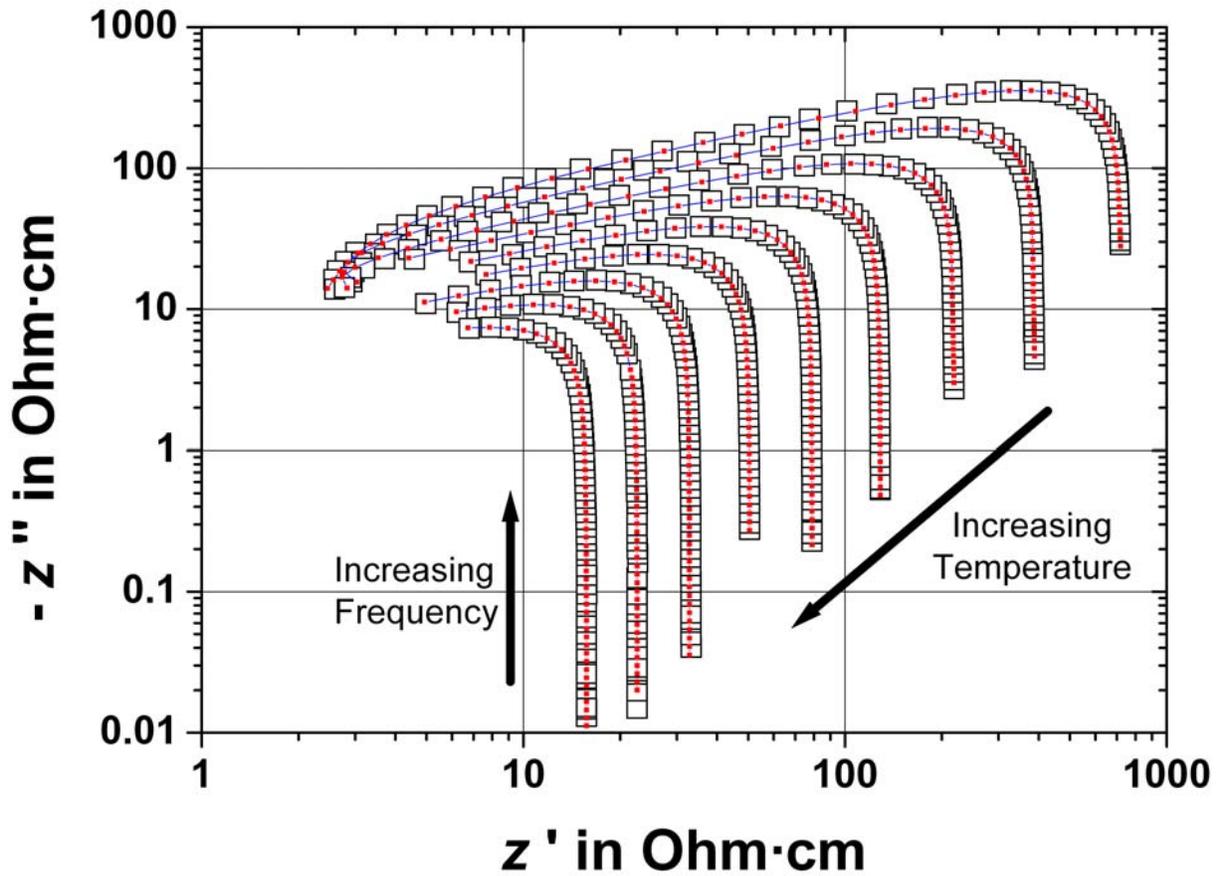

**Fig 3 (a)** Specific capacitance c' vs frequency $f$ : data (□) and fits using the model in Fig 2 with R1-CPE1 with fixed $n1 = 0.75$ (■) and free $n1$ (■) , and for a circuit of one single Debye RC element (■) ; data was taken at 120ºC **(b)** Nyquist loci of –$z''$ vs $z'$, □ data and fit with fix $n1$ (■) using the equivalent circuit model in Fig 2, data taken at 60ºC - 220ºC in 20ºC steps

The $c'$ and $\rho$ vs $T$ curves obtained from the fits are shown in Fig 4. Fitting errors are shown, except for R2 and C2, where the errors were too low to be displayed. Transitional behaviour was found in C1 vs $T$ around 100ºC. The Fig 4(a) inset demonstrates the transition on a linear scale.

Different fixed $n1$ values were used for $T \leq 100°C$ and for $T \geq 120°C$ ($n1_{lowT} = 1$; $n1_{highT} = 0.75$) to optimise the fit, which could potentially be the main factor inducing such transitional behaviour. It was attempted to fit the data at high and low $T$ with converging $n1$ values around 100°C, i.e. $n1_{lowT} = 0.95$; $n1_{highT} = 0.8$ // $n1_{lowT} = 0.9$; $n1_{highT} = 0.85$ // $n1_{lowT} = n1_{highT} = 0.875$. It was expected that the C1 vs $T$ curve would level out if the transitional behaviour was induced solely by $n1$. The opposite trend is shown and the indications for a transition are significant.

(a)

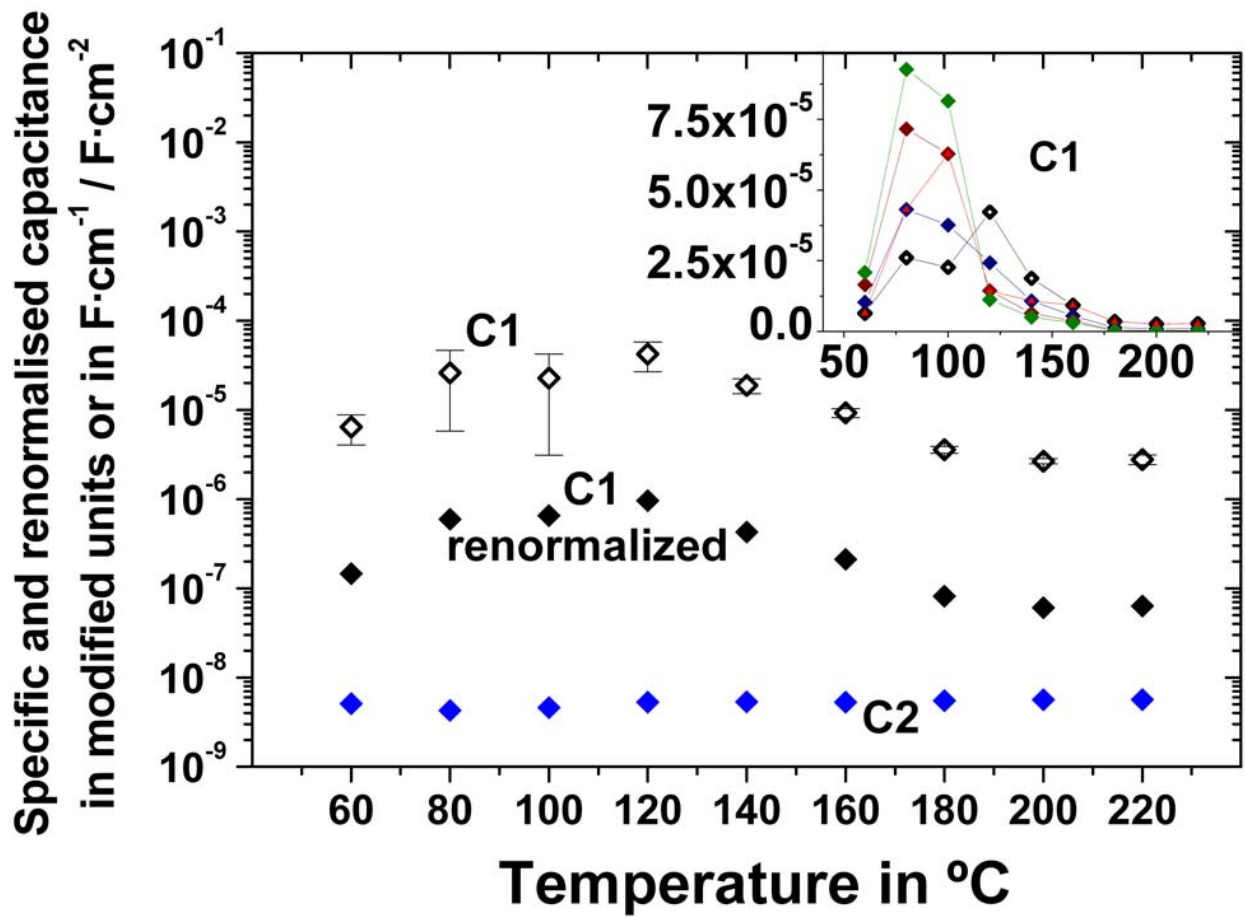

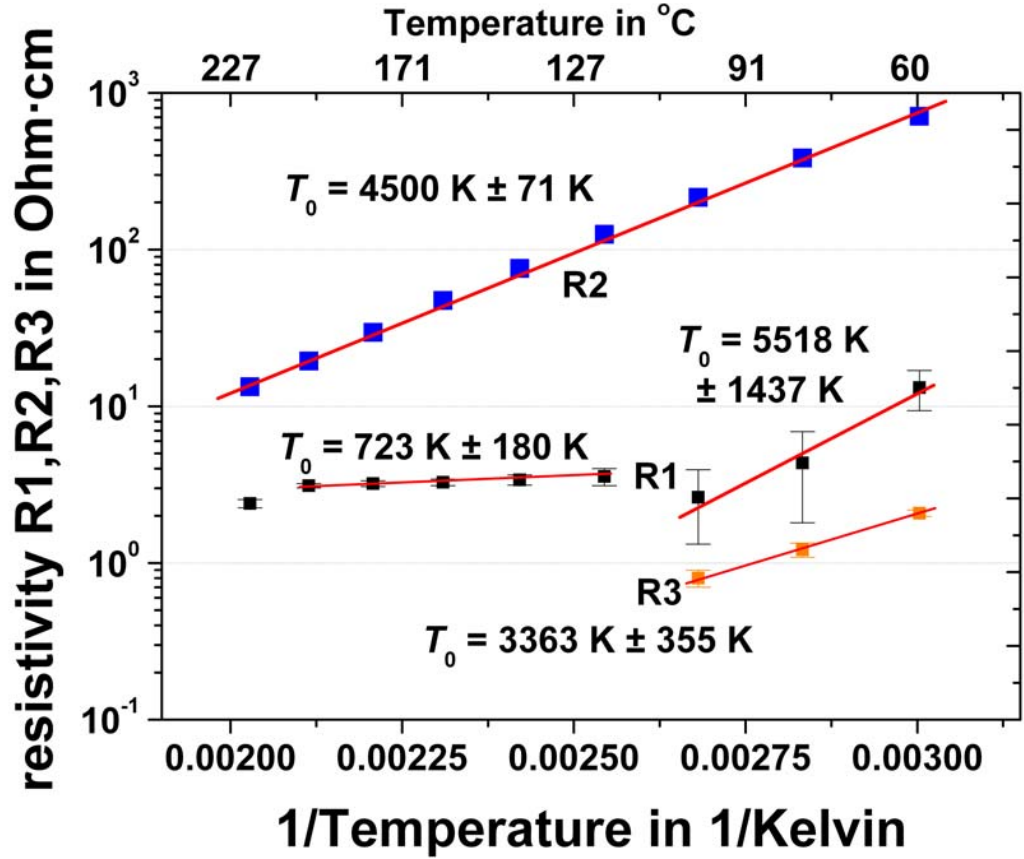

**Fig 4 (a)** C1 (◊), C1 renormalized (♦) and C2 (♦) vs *T*; for *T* ≤ 100ºC: C1 is given in F·cm$^{-1}$ and renormalized to F·cm$^{-2}$, *T* ≥ 120ºC: C1 in F·s$^{-0.25}$·cm$^{-1}$ and renormalized to F·s$^{-0.25}$·cm$^{-2}$; C2 is given in F·cm$^{-1}$; errors are indicated for C1, errors in C2 were too low to be displayed, **Inset:** C1 vs *T* (◊) with $n1_{lowT} = 1$ and $n1_{highT} = 0.75$; for converging $n1$: $n1_{lowT} = 0.95$; $n1_{highT} = 0.8$ (♦)// $n1_{lowT} = 0.9$; $n1_{highT} = 0.85$ (♦)// $n1_{lowT} = n1_{highT} = 0.875$(♦); and for gradual changes (▲): $n1(60ºC) = 1$; $n1(80ºC) = 0.95$… $n1(T \geq 160ºC) = 0.75$; **(b)** Resistivity vs temperature dependence of R1, R2 and R3; the characteristic temperatures $T_0$ were determined from $\ln(\rho/T)$ vs $1/T$ plots to be 723K, 5518K, 4500K and 3363K, which correspond to activation energies of 0.06eV, 0.48eV, 0.39eV and 0.29eV respectively, assuming NNH according to eq.(1)

In addition, data was fitted by gradually decreasing $n1$, i.e. $n1(60°C) = 1$; $n1(80°C) = 0.95$ etc., with a step size of 0.05 and constant $n1 = 0.75$ for $T \geq 160°C$. This was believed to be the physically most reasonable approach and indeed resulted in a smooth C1 vs $T$ curve with an emphasized transition. For data fitting the model with constant high$T$ and low$T$ $n1$ proved to be superior, because it allowed better comparison of the capacitances obtained from CPEs with identical $n$-parameters and showed higher consistency in the $\rho$ vs $T$ curves. The capacitance C1 was renormalized in order to compare to the classification scheme proposed by Irvine et al.[6] In bulk material measurement geometries the contact area is often identical to the current cross section $A$ and the interface capacitance can be scaled by $A$. This is not the case in the films presented here, which were deposited on an insulating $Al_2O_3$ substrate and the in-plane impedance was measured with a top-top arrangement of contacts and spring loaded drop down probes. C1 was renormalized to the contact area, which resulted in $c'$ values typical for an interface. For $T \geq 100°C$, C1 in Fig 4(a) is displayed in modified units of $F \cdot s^{n-1} \cdot cm^{-1}$. It was attempted to convert C1 to units of $F \cdot cm^{-1}$ according to the method proposed by Hsu & Mansfeld.[21] This lead to C1 vs $T$ behaviour qualitatively similar as shown in Fig 4(a), but to an essential reduction in C1 by a factor of up to ~ 20 for $T \geq 120°C$. These values were believed to be unreasonable and it is suggested that the physical meaning of CPE behaviour for interface contributions may be ambiguous, and conversion of the CPE1 parameters according to ref. [21] may not be valid.

The C2 vs $T$ dependence in Fig 4a revealed a capacitance approximately constant with $T$, which is expected for an intrinsic contribution in absence of any phase transition. C2 was normalised to $g$ ($=A/d$) and was in the range of a grain boundary contribution. The $\rho$-$T$ behaviour of R2 indicated close agreement with the dc $\rho$-$T$ dependence (a detailed description is given below), which leads to the conclusion that the glass phase incorporated into the thick films did not have an effect in the intermediate frequency regime dominated by R2-C2. Thus, the identification of a grain boundary effect was believed to be reliable and effects from the glass phase negligible, which was further supported by a recent study on the surface microstructure of such films, revealing direct contact of grains constituting a charge carrier percolation path through the sample.[22]

The C3 vs $T$ behaviour is not shown, because the errors were meaninglessly high and C3 may contain perceptible high frequency noise contributions. The $\rho$-$T$ dependencies shown in Fig 4(b) suggested thermally activated transport behaviour in all three contributions, where the slopes of the $\ln(\rho/T)$ vs $1/T$ plots yielded the characteristic temperatures $T_0$ and activation energies for a NNH process according to eq.1. The contact resistance R1 was relatively small, which may be a result of the spring loaded probes used and/or the intimate contact of electrode material and film due to the thermal evaporation process employed for Al/Ag contact deposition. Transitional behaviour of the interface at ~ 100ºC was indicated in the R1 vs $T$ curve by a change in activation energy, although the errors in R1 below 100ºC are considerably high. Close inspection of Fig 4(b) shows that the R2 curve is slightly bended, typical for VRH. An estimate for the VRH parameter $p$ in eq.1 was obtained as the negative slope of a plot of $\ln W$ vs $\ln T$. The $\ln W$ method is based on differentiated $\ln(\rho)$ vs $1/T$ data and has been described in a previous publication.[14] In Fig 5 the $\ln W$ vs $\ln T$ plots are shown for R2 and for R1+R2+R3 vs $T$.

Good linearity of both curves confirmed the validity of eq.1 for VRH. The close agreement in $p$ indicated that the hopping mechanism of the R2-C2 element and the macroscopic behaviour varied only marginally. A $p$-value of 0.6 for R2 is in good agreement with the value obtained from the single Debye semicircle fitting procedure described in ref. [20] ($p$ = 0.6). $T_0$ was determined to be $5.44 \cdot 10^4$ K for $p = 0.6$. For R1+R2+R3, $p$ was reasonably close to the value reported from dc measurements for polycrystalline bulk samples ($p = 0.65$).[14] Assuming $p = 0.65$, $T_0$ was determined to be $2.3 \cdot 10^4$ K, which implies a weaker temperature dependence of the films compared to pellets ($T_0 = 3.1 \cdot 10^4$ K).

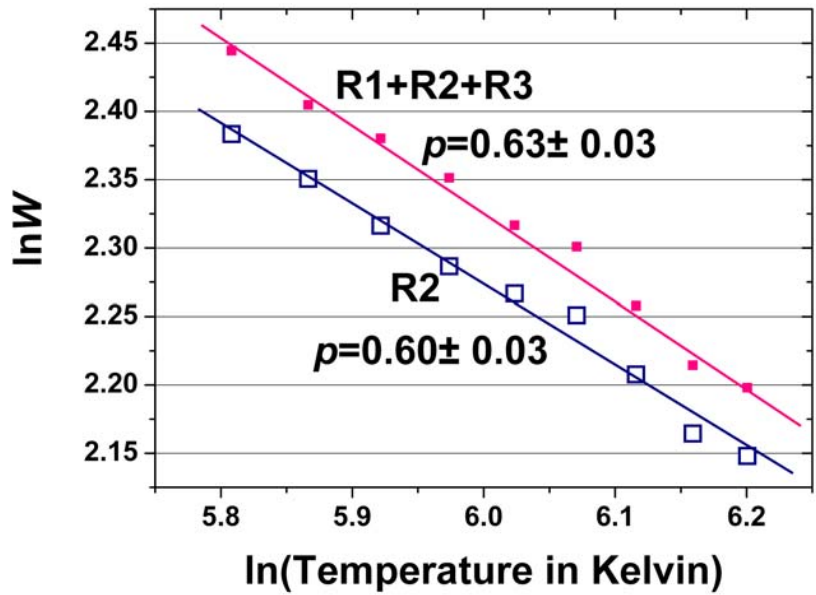

**Fig 5** $\ln W$ vs $\ln T$ plots for R2 (□) and (R1+R2+R3) (■); the (R1+R2+R3) curve is displaced by 0.05 up the $\ln W$ axis

The close agreement of $p$ in thick films and in glass free pellets further supports the claim of a negligible influence of the glass phase in the intermediate frequency regime. It also suggests that charge transport in polycrystalline thick films varies only quantitatively from bulk material pellets. In the most likely scenario, R1-CPE1 may describe an electrode-sample interface effect,

R2-CPE2 the dominating grain boundary conduction and R3-CPE3 could be an indication for a small bulk contribution and/or the insulating glass phase at high frequency.

## 4. Conclusions

In conclusion, it has been shown that the complex impedance in thick film nickel manganate is dominated by one relaxation. At the low and high frequency ends additional contributions were described by R-CPE$_C$ elements and the composite character of the film response was demonstrated. The low frequency relaxation showed a capacitance typical for an electrode interface, and indications for transitional behaviour at ~ 100ºC. The electrode resistivity did not affect the macroscopic small polaron hopping mechanism significantly, which would be beneficial for the fabrication of reproducible devices for thermistor applications. In the literature, different models for the $\rho$-$T$ behaviour of polycrystalline spinel manganate ceramics have been suggested based on dc measurements,[23,24] but the composite nature of such resistivity has not been discussed so far. The main relaxation demonstrated here originates most likely from grain boundary conduction and is believed to represent a realistic description of intrinsic NTCR behaviour of polycrystalline thermistor thick film ceramics.

## 5. Methods

Thick film NTCR spinel manganate thermistor devices are interesting for a wide range of application and can be produced by screen printing.[22,25] Such films exhibit dense grain packing and low surface porosity. A description of the experimental setup for temperature dependent ac impedance spectroscopy measurements between 60°C – 220°C using an HP 4192A LF Impedance Analyzer at frequencies of 5Hz – 6MHz can be found elsewhere.[20] The impedance analyzer was operated with an alternating voltage signal of 3V amplitude, superimposing a constant dc bias of 30V to reduce noise. Initially, open and short circuit measurements were carried out in order to correct the data taken from the sample by subtracting parasitic contributions. The data fitting was performed using commercial software (Z-View), where for each experimental data point an optimised value was obtained from the model. The impedance data collected between 60°C – 220°C in 20°C steps were normalized by multiplying by the geometrical factor $g$ or the contact area, which yielded the respective specific capacitance and the resistivity. The geometrical factor was estimated from the sample geometry, and the resistivity and specific capacitance may be regarded as an approximated value only. The data was cut at low and high frequency ends where appropriate in order to not display and fit sample response dominated by noise.

**Acknowledgments**

The authors wish to thank Michael Petty for allowing use of the ac impedance spectroscopy facility. Thanks to Finlay Morrison and Ian Terry for useful discussions concerning impedance spectroscopy data analysis. Thanks to Andreas Roosen and Alfons Stiegelschmitt for the guidance provided in developing screen-printing procedures.

# Supporting Materials

## Parameters obtained from the fit at different temperatures

Table 1 a) Fit parameters R1, *n*1, C1, R2, *n*2, C2

| Temp. in °C | R1 in Ohm cm | *n*1 fixed | C1 in F cm$^{-1}$ & F s$^{-0.25}$ cm$^{-1}$ | R2 in Ohm cm | *n*2 fixed | C2 in F cm$^{-1}$ |
|---|---|---|---|---|---|---|
| 60  | 13.2± 3.8   | 1    | 6.44E-6± 2.4E-6 | 709.4± 1.5 | 1 | 5.10E-9± 7.8E-12 |
| 80  | 4.35± 2.5   | 1    | 2.61E-5± 2.0E-5 | 383.8± 1.0 | 1 | 4.27E-9± 1.1E-11 |
| 100 | 2.63± 1.3   | 1    | 2.87E-5± 2.0E-5 | 215.6± 0.5 | 1 | 4.59E-9± 1.2E-11 |
| 120 | 3.56±0.44   | 0.75 | 4.22E-5± 1.5E-5 | 125.4± 0.5 | 1 | 5.29E-9± 1.3E-11 |
| 140 | 3.39±0.25   | 0.75 | 1.88E-5± 3.6E-6 | 75.7± 0.3  | 1 | 5.32E-9± 1.4E-11 |
| 160 | 3.27±0.16   | 0.75 | 9.27E-6± 1.1E-6 | 47.3± 0.2  | 1 | 5.30E-9± 1.8E-11 |
| 180 | 3.21±0.14   | 0.75 | 3.58E-6± 3.2E-7 | 29.7± 0.1  | 1 | 5.51E-9± 4.2E-11 |
| 200 | 3.11±0.10   | 0.75 | 2.66E-6± 1.8E-7 | 19.5± 0.1  | 1 | 5.63E-9± 4.6E-11 |
| 220 | 2.40±0.14   | 0.75 | 2.78E-6± 3.4E-7 | 13.33± 0.1 | 1 | 5.66E-9± 8.8E-11 |

Table 1b) Fit parameters R3, *n*3, C2

| Temp. in °C | R3 in Ohm cm | *n*3 free | C2 F s$^{-0.25}$ cm$^{-1}$ |
|---|---|---|---|
| 60  | 2.08± 0.10 | 1.33± 0.10 | *5.07E-11± 8.2E-11* |
| 80  | 1.22± 0.13 | *2.06± 0.05* | 7.01E-16± 6.0E-16 |
| 100 | 0.80± 0.1  | *2.12± 0.04* | 4.69E-16± 3.9E-16 |